\documentclass[11pt]{article}
\usepackage{amssymb}
\usepackage{amstex}
\input{psfig.sty}
\begin{document}
\pagestyle{plain}
\hsize = 6.5 in
\vsize = 9.0 in
\hoffset = -0.75 in
\voffset = -1.00 in
\def \rvX{{\bf X}}
\def \rvY{{\bf Y}}
\def \rvz{{\bf Z}}
\def \rvT{{\bf T}}
\def \rvD{\hbox{\boldmath $\Delta$}} 
\def \rt{\hbox{\boldmath $\xi$}} 
\def \rphi{\hbox{\boldmath $\phi$}} 
\baselineskip=0.23in

\title{\Large{\bf A Stochastic Analysis of a Brownian Ratchet Model for\\
Actin-Based Motility and Integrate-and-Firing Neurons}}
\author{\normalsize{Hong Qian} \\
\\
\normalsize{Department of Applied Mathematics, University of Washington,
Seattle, WA 98195}}
\date{\normalsize{\today}}

\maketitle

\vskip 0.3cm \noindent
{\bf Abstract.}  
In recent single-particle tracking (SPT) measurements on {\it Listeria 
monocytogenes} motility {\em in vitro}, the actin-based stochastic 
dynamics of the bacterium movement is analyzed statistically (Kuo and 
McGrath, 2000).  The mean-square displacement (MSD) of the detrended
trajectory exhibit a linear behavior; it has been suggested that 
a corresponding analysis for the Brownian ratchet model (Peskin, 
Odell, \& Oster, 1993) leads to a non-monotonic MSD.  A simplified
version of the Brownian ratchet, when its motion is limited by the 
bacterium movement, is proposed and analyzed stochastically.  
Analytical results for the simple model are obtained
and statistical data analysis is investigated.  The MSD of 
the stochastic bacterium movement is a quadratic function
while the MSD for the detrended trajectory is shown to be linear.  
The mean velocity and effective diffusion constant of the propelled 
bacterium in the long-time limit, and the short-time relaxation 
are obtained from the MSD analysis.  The MSD of the gap between 
actin and the bacterium exhibits an oscillatory behavior 
when there is a large resistant force from the 
bacterium.  The stochastic model for actin-based motility 
is also mathematically equivalent to a model for 
integrate-and-firing neurons.  Hence our mathematical results 
have applications in other biological problems.  For comparison,
a continuous formalism of the BR model with great analytical 
simplicity is also studied.

\centerline{\line(1,0){230}}

  

\vskip 1.0cm \noindent
{\bf 1. Introduction}
\vskip 0.3cm \noindent

	Actin polymerization plays an important role in nonmuscle
cell mechanics, motility, and functions (Pollard et al, 2000;
Pantaloni et al., 2000).  In recent years, quantitative analyses 
of the molecular mechanism for actin-based motility are made 
possible by both laboratory experiments on {\it Listeria monocytogenes} 
(see van Oudenaarden and Theriot, 1999, and the references 
cited within) and  a series of insightful mathematical models
(Hill, 1981, 1987, Peskin et al., 1993, Mogilner and Oster 1996).  
The interaction between experimental observations and theoretical 
ideas has generated exciting research in biophysics and 
mathematical biology. 

	Following the seminal work of Peskin et al. (1993),
a sizable literature now exists on 
mathematical models and analyses of the polymerization-based 
motility, known as Brownian ratchet (BR).  Even though the 
original model on fluctuations is clearly a probabilistic one, 
it was cast mathematically in terms of the difference and
differential equations with only a minimal stochastic interpretation.  
In the subsequent development, this stochastic nature of the model 
often has been obscured.  In experimental laboratories, on the 
other hand, researchers often use Monte Carlo simulations
to model the biological problem, partly because the data 
are inevitably stochastic.

	This situation has prevented a truly quantitative
understanding of the actin-based motility and a closer 
interaction between the experimental measurements and 
mathematical modeling.  In a recent experiment, Kuo and McGrath 
(2000) used the highly sensitive single-particle tracking 
(SPT) methodology to measure the stochastic movement of 
{\it L. monocytogenes} propelled by actin polymerization.
The seemingly random data are then analyzed statistically in 
terms of the mean-square displacement (MSD).  The exquisite
data with nanometre precision reveals the discrete steps in
the bacteria movement, presumably due to the actin 
polymerization, one G-actin monomer at a time. 

	The stochastic nature of the BR, and the statistical
treatment employed in experimental data analyses,
necessitate a mathematical analysis of the BR model in
fully stochastic terms.  This is the main objective of the
present work.  Furthermore, Kuo and McGrath (2000) suggested that
the BR movement, after detrending, exhibits a non-monotonic MSD.  
We shall investigate these practical issues as well.  The 
significance of the stochastic interpretation is that one needs
only to think about a single BR, and can derive theoretical MSD 
to compare with experiments. 

	In order to clearly present the stochastic approach
to the BR, we study only a special, but relevant, case of the 
generic BR model proposed by Peskin et al (1993).  This restriction
makes the model easily analyzed analytically.  Interestingly, the 
mathematical model is also identical to one for integrate-and-firing
neuron proposed many years ago by Gerstein and Mandelbrot (1964).
In recent years, integrate-and-firing model has become one of the
essential components in neural modeling (Hopfield and Herz, 1995).
The stochastic model is quite basic; therefore we expect that our 
mathematical results also have applications in other branches of 
mathematical biology.  The fractal nature of such model has also
been discussed recently (Qian et al., 1999).

	All the mathematical background on stochastic processes 
used in this work can be found in the excellent text by Taylor 
and Karlin (1998).  To help the readers who are not familiar with 
some of the stochastic mathematics, italic font is used for the 
key words when they first appear in the paper.

\vskip 0.5cm \noindent
{\bf 2. Stochastic Formulation of a Brownian Ratchet Model}
\vskip 0.3cm \noindent

	(i) We consider an F-actin polymerizes in a 1-dimensional
fashion with the rate of monomer addition $\alpha$ and the rate 
of depolymerization $\beta$.  $\alpha$ is a pseudo-first order
rate constant which is proportional to the G-actin monomer
concentration.  Each G-actin monomer has a size of $\delta$.
Hence the actin polymerization is modeled as a continuous-time 
random walk (Hill, 1987).  We shall take the growing direction 
as positive, and denote the position of the tip of the actin 
filament by $\rvX(t)$ which is a stochastic process taking 
discrete values $k\delta$, where $k$ is an integer.

	(ii) We assume that a bacterium is, in the front of the 
growing actin filament, located at $\rvY(t)$: $\rvX(t)\le\rvY(t)$.  
The bacterium has an intrinsic diffusion constant $D_b$, 
and experiencing (or exerting) a resistant force $F$ in the 
direction against the actin polymerization.  In the absence
of the actin filament, the bacterium movement is a Brownian
motion with a constant drift rate $-F/\eta_b$.  Since a 
bacterium is a living organism, the $D_b$ and 
the $\eta_b$ are not necessarily related by the Einstein 
relation $\eta_bD_b= k_BT$ for inert equilibrium objects.

	(iii) The F-actin and the bacterium interact
only when they encounter: $\rvX(t)=\rvY(t)$.  The
actin filament, however, can not penetrate the bacteria
wall.  Therefore, the motion of the bacterium and the actin
polymerization are coupled via a reflecting boundary
condition at $\rvX(t)=\rvY(t)$.  

	(i)-(iii) are the basic assumptions of the generic
BR model first proposed by Peskin et al. (1993).  In the 
present work, we shall further assume that (iv) the $\alpha$
is sufficiently large and (v) $\beta \approx 0$.  Therefore, 
whenever the gap $\rvD(t)\triangleq \rvY(t)-\rvX(t)$
= $\delta$, the gap will be immediately filled by a 
G-actin monomer, and the polymer does not depolymerize.    
These two assumptions correspond to a rapid polymerization
condition under which the bacteria movement is the rate-limiting 
process in the overall kinetics. 

	Fig. 1 shows the basic, stochastic behavior of 
$\rvX(t)$, $\rvY(t)$, and $\rvD(t)$.  Kuo and McGrath (2000)
also introduced a detrended $\rvY(t)$.  Let $v$ be the 
mean velocity of the bacterium movement $\rvY(t)$, then the 
detrend $\hat{\rvY}(t)$ is defined as 
$\hat{\rvY}(t)\triangleq\rvY(t)-vt$. 

	Let $\rt_k$ be the time for incorporating the k$th$ 
G-actin monomer.  Then at time $\rt_k$, 
$\rvX(\rt_k)=\rvY(\rt_k)=k\delta$.  When $t>\rt_k$, 
$\rvY(t)$ follows a Brownian motion with diffusion 
constant $D$, drift rate $-F/\eta_b$, and reflecting boundary 
at $k\delta$.  $\rvY(t)$ moves stochastically and when it
reaches $(k+1)\delta$, denoted the time by $\rt_{k+1}$, 
the (k+1)$th$ G-actin monomer is incorporated.  Then 
the process repeats.  The waiting time for the next 
monomer to be incorporated is a random variable, we
shall denote it by $\rvT$: $\rt_{k+1}=\rt_k+\rvT$.  This is 
our stochastic formalism for the BR model.  Our analysis
focuses on the stochastic properties of the random variable
$\rvT$.  

	Fig. 2 shows the mean-square displacement (MSD)
of the stochastic data in Fig. 1.  The MSD for a stochastic
processes $\rvX(t)$ is defined as 
\begin{equation}
        MSD(\tau) = E\left[(\rvX(\tau+t)-\rvX(t))^2\right]
\label{msd}
\end{equation}
which is a powerful analytical tool for analyzing stochastic
processes with {\it independent increments} or
{\it stationarity}.  The $E[\hdots]$ in 
Eq. \ref{msd} denotes the expectation of random variables.  
For a stochastic process with independent increments, 
MSD$(\tau)$ is further simplified into 
$E\left[(\rvX(\tau)-\rvX(0))^2\right]$.
In the case of a stationary process, its MSD is directly 
related to the correlation function:
\begin{equation}
        E[\rvX(\tau)\rvX(0)] = E[\rvX^2]
                             - \frac{1}{2}MSD(\tau).
\label{msdcf}
\end{equation}

	The significance of MSD is that it can be obtained 
through a statistical analysis of stochastic experimental 
data (Qian et al., 1991).  It is the essential link
between the experimental measurements on fluctuations and 
stochastic mathematical models.  For an experimental time 
series $\{x_n|0\le n\le N\}$, the MSD is defined as:
\begin{equation}
          MSD(m) = \frac{1}{N-m+1}\sum_{k=0}^{N-m} 
			\left(x_{k+m}-x_k\right)^2.
\label{msd2}
\end{equation}
The statistical relation between the experimentally determined
MSD in Eq. \ref{msd2} and the theoretical MSD in Eq. \ref{msd}
can be found in the paper by Qian et al. (1991).

\vskip 0.5cm \noindent
{\bf 3. Basic Properties of the Model: Analytical Results}
\vskip 0.3cm \noindent

{\bf\em Mean Waiting Time and Waiting Time Distribution.}  
The time interval $\rvT$ between the repeated incorporation of 
successive actin monomer is the {\it exit time} of a diffusion 
process. By exit time $\rvT_z$, we mean the time a Brownian 
particle takes to reach $\delta$ the first time, starting at 
$z$ $(0\le z\le \delta)$.  Clearly $\rvT_z$ is a random variable;  
its expectation $T(z)$ = $E[\rvT_z]$, known as {\it mean first 
passage time}, is the solution to the differential equation 
(Taylor and Karlin, 1998)
\begin{equation}
        D_bT''_{zz}-(F/\eta_b)T'_z = -1
\label{mfpt}
\end{equation}
with boundary conditions $T'_z(0)=0$ and $T(\delta)=0$.  Hence
\begin{equation}
          T(z) = \frac{\eta_b^2 D_b}{F^2}\left(e^{F\delta/\eta_bD_b}
	- e^{Fz/\eta_bD_b}\right)+\frac{\eta_b(z-\delta)}{F}.
\end{equation}
Therefore,
\begin{equation}
       E[\rvT] = T(0) = \left(\frac{\delta^2}{D_b}\right)
			\frac{e^{\omega}-1-\omega}{\omega^2},
\label{Texp}
\end{equation}
where $\omega=F\delta/(\eta_bD_b)$ is the nondimensionalized resistant
force. 

The probability density function $f_{\rvT_z}(t)$ for the waiting time, 
$\rvT_z$ can be obtained in terms of its Laplace transform, also 
known as the {\it characteristic function} of the random variable 
$\rvT_z$, $Q_{\rvT}(z,\nu)$ = $\int_0^{\infty}f_{\rvT_z}(t)e^{-\nu t}dt$
which satisfies the following differential equation (Weiss, 1966)
\begin{equation}
    D_b\frac{\partial^2Q_{\rvT}(z,\nu)}{\partial z^2}
         -\frac{F}{\eta_b}\frac{\partial Q_{\rvT}(z,\nu)}{\partial z} 
			= \nu Q_{\rvT}(z,\nu)
\label{dfpt}
\end{equation}
with boundary condition $\partial Q_{\rvT}(0,\nu)/\partial z=0$ and 
$Q_{\rvT}(\delta,\nu)=1$.  Note Eq. \ref{mfpt} is a special case 
of Eq. \ref{dfpt} for $T(z)=-\partial Q_{\rvT}(z,0)/\partial\nu$. 

	Eq. \ref{dfpt} can be analytically solved:
\begin{equation}
     Q_{\rvT}(0,\nu) = \frac{\lambda_- - \lambda_+}
     {\lambda_-e^{\lambda_+\delta}-\lambda_+e^{\lambda_-\delta}}
\label{mgf}
\end{equation} 
where
\[          \lambda_{\pm} = \frac{\omega}{2\delta}\pm \sqrt{
		\left(\frac{\omega}{2\delta}\right)^2+\frac{\nu}{D_b}}.
\]
Therefore, the variance in the waiting time
\begin{equation}
	Var[\rvT] = \left(\frac{\delta^4}{D_b^2}\right)
	\frac{3e^{2\omega}-(10\omega-6)e^{\omega}+\omega^2
		-2\omega-9}{\omega^4}.
\label{Tvar}
\end{equation}
If there is no resistant force from the bacteria, $\omega=0$ and 
we have a simple expression
\begin{equation}
   Q_{\rvT}(0,\nu) = \left(\cosh\sqrt{\delta^2\nu/D_b}\right)^{-1}.
\end{equation}

\vskip 0.3cm
{\bf\em Renewal Processes, The Statistical Properties of $\rvX(t)$
and $\rvY(t)$.}
With $\rvT$ as the waiting time, the tip of the rapid growing
actin filament, $\rvX(t)$, is a {\it renewal process}.  There is
a large literature on this subject.  The most relevant result to 
our model is the {\it elementary renewal theorem} for large $t$ 
\begin{equation}
         E[\rvX(t)] \approx \frac{\delta}{E[\rvT]}t
\end{equation}
Therefore as a renewal process, a BR executes successive steps 
with size $\delta$ and average time $E[\rvT]$.  The mean velocity 
of the BR, thus, is 
\begin{equation}
   v = \frac{\delta}{E[\rvT]} = \left(\frac{D_b}{\delta}\right)
			\frac{\omega^2}{e^{\omega}-1-\omega}.
\label{BRv}
\end{equation}
This result is in agreement with that of Peskin et al. 
(1993).  

	Furthermore from the theory of renewal process 
(Taylor and Karlin, 1998) 
\begin{equation}
     Var[\rvX(t)] \approx \frac{\delta^2Var[\rvT]}{E^3[\rvT]}t
	\triangleq \sigma^2t,
\end{equation}
where, according to Eqs. \ref{Texp} and \ref{Tvar},
\begin{equation}
     \sigma^2 = \frac{3e^{2\omega}-(10\omega-6)e^{\omega}
	+\omega^2-2\omega-9}{(e^{\omega}-1-\omega)^3}\omega^2D_b.
\label{sig}
\end{equation}
The MSD for $\rvX(t)$, therefore, is
\begin{equation}
       E\left[(\rvX(t)-\rvX(0))^2\right]\approx \sigma^2t + (vt)^2 
\end{equation}
which is a quadratic function of $t$.  The expression for $\sigma^2$ 
is a new result of the present work, which is comparable with 
experimental data.  Fig. 3 shows the dependence of $v$ and 
$\sigma^2$ as functions of $\omega=F\delta/\eta_bD_b$, the resistant 
force from the bacterium. If $\omega = 0$, then $v=2D_b/\delta$ and
$\sigma^2$ = $14D_b/3$.  

	Since $\rvY(t)-\rvX(t)<\delta$ while both increase linearly
with $t$, for large $t$ $\rvY(t) \approx \rvX(t)$ with an error
less than $\delta$, the size of a single actin monomer. 
Strictly speaking, the $\rvY(t)$ is not a stochastic process 
with independent increments.  However, the error involved, 
again, is only on the order of the size of a single G-actin.  
To understand the statistical correlation of $\rvY(t)$ within 
each ``step'', see the section below on the gap.

\vskip 0.3cm

{\bf\em Detrend $\rvY(t)$ and Its Statistical Properties.}
In the recent experimental work (Kuo and McGrath, 2000), the 
detrended $\rvY(t)$ has also been reported, which can be defined 
as $\hat{\rvY}(t) \triangleq \rvY(t) -vt$, where $v$ is the mean
velocity. For $n\delta \le \rvY(t) \le (n+1)\delta$, 
\begin{equation}
 \hat{\rvY}(t) \triangleq      
            \rvY(t) -vt = \rvY(t)-\rvY(\rt_n)
		+n\delta-v\rt_n -v(t-\rt_n)
                     = \rvY(\tau)-v\tau+n\delta-v\rt_n
\end{equation}
in which random variable $\rt_n$ is the time for $\rvY(t)$ to reach 
$n\delta$ the first time, $\tau = t-\rt_n$, and $v$ is given in 
Eq. \ref{BRv}.  Hence, $0\le \tau\le \rt_{n+1}-\rt_n$ = $\rt_1$ 
= $\rvT$, with its expectation, variance, and characteristic
function given in Eqs. \ref{Texp}, \ref{Tvar}, and \ref{mgf},
respectively. 

	The statistical properties of $\hat{\rvY}(t)$ are 
readily to be calculated:
\begin{equation}
       E\left[\hat{\rvY}(t)\right] = E\left(\rvY(\tau)\right]-v\tau
	\approx 0,
\end{equation}
\begin{equation}
       Var\left[\hat{\rvY}(t)\right] = Var\left[\rvY(\tau)\right]
		+nvVar\left[\rvT\right] \approx \sigma^2t.
\label{vdt} 
\end{equation}
Thus, we see that the detrend $\hat{\rvY}(t)$ does not become
stationary with increasing time.  While its expectation is
zero, its variance increases linear with the time $t$, the
epitome of a symmetric random movement.  The parameter 
$\frac{\sigma^2}{2}$ is the effective diffusion constant of 
the BR.

\vskip 0.3cm \noindent

{\bf\em Statistical Properties of the Gap.}  The gap between the 
bacteria, $\rvY(t)$, and the tip of the actin filament $\rvX(t)$,
$\rvD(t)\triangleq \rvY(t)-\rvX(t)$ behaves completely 
different from the detrend $\hat{\rvY}(t)$.   It reaches 
asymptotically to stationarity. 

	We can provide a reasonable estimation for the 
relaxation time for the gap to reach its stationarity 
from the largest nonzero eigenvalues ($\mu$) of the 
diffusion operator  
\begin{equation}
    \left(D_b\frac{d^2}{dx^2}+\frac{F}{\eta_b}\frac{d}{dx}\right) 
		u(x) = \mu u(x) 
\label{eigen}
\end{equation}
under the boundary condition $D_bu'(0)+(F/\eta_b)u(0)$ =
$u(\delta)$ = 0.  See Appendix for details.  All the eigenvalues 
are real and $\le 0$, $\mu(z)$ = $-\frac{D_b}{4\delta^2}(z^2+\omega^2)$, 
where the $z$ are the roots of the transcendental equation
$\cos z$ = $\frac{\omega^2-z^2}{\omega^2+z^2}$.  Fig. 4 suggests that 
the largest eigenvalue corresponds to the exit time which increases 
with the resistant force.  For resistant force 
$F\gg 2\eta_bD_b/\delta$, there is a separation between 
the time scale for the exit and the time scale for 
establishing a quasi-stationary distribution for $\rvD(t)$ (Appendix).  
Fig. 5 shows the MSD for $\rvD(t)$, which is directly related to 
the correlation function for the stationary process (Eq. \ref{msdcf}).  
After normalized by $2Var[\rvD]$, the MSD are approximately 
the same for $0\le\omega\le 6$.  The correlation time decreases with 
$\omega$ for $\omega$ = 2, 4, 6, and 12 (i.e., $p$ =
0.55, 0.6, 0.65, and 0.8), corresponding to the second largest 
eigenvalue in Fig. 4,

	When there is a large resistant force $F$, the exit time 
$\rvT$ has a small relative variance (Eqs. \ref{Tvar}, and \ref{Texp})
and the exit becomes an event with sufficient regularity.  This is 
reflected in the oscillation of the MSD in Fig. 5.

\vskip 0.5cm \noindent
{\bf 4. An Analytical Analysis of a Continuous Stochastic Formalism of BR}
\vskip 0.3cm \noindent

	We can replace the assumptions (iv) and (v) in 
the Section 2 with a continuous model for the discrete 
polymerization.  In other word, we approximate the random
walk by a diffusion with diffusion constant and drift
rate (Feller, 1957; Hill, 1987):
\begin{equation}
           D_a=(\alpha+\beta)\delta^2/2,   \hspace{0.5cm}
           V_a=(\alpha-\beta)\delta.
\label{corresp}
\end{equation}
where $\beta$ and $\alpha$ are first-order and pseudo-first-order 
rate constants for the depolymerization and polymerization, $\delta$
is the size of a G-actin monomer.  The dynamic equation governing 
the probability density function (pdf) $P_{\rvX}(x,t)$ for the 
stochastic processes $\rvX(t)$ is:
\begin{equation}
  \frac{\partial P_{\rvX}(x,t)}{\partial t} =
     D_a\frac{\partial^2 P_{\rvX}(x,t)}{\partial x^2} - 
     V_a\frac{\partial P_{\rvX}(x,t)}{\partial x}     
\label{xdyn}
\end{equation}
where $P_{\rvX}(x,t)$ has the probabilistic meaning of
$P_{\rvX}(x) dx = Prob\{x\le \rvX < x+dx\}$.  Similarly, 
the dynamical equation for Brownian motion of the bacterium
$\rvY$ with resisting force $F$ is, as before,
\begin{equation}
  \frac{\partial P_{\rvY}(y,t)}{\partial t} =
     D_b\frac{\partial^2 P_{\rvY}(y,t)}{\partial y^2} + 
     \frac{F}{\eta_b}\frac{\partial P_{\rvY}(y,t)}{\partial y}.     
\label{ydyn}
\end{equation}
These two equations are coupled since $\rvX\le\rvY$.  
We call Eqs \ref{corresp}, \ref{xdyn}, and \ref{ydyn}
the continuous formalism for the BR.  It represents a 
two-dimensional diffusion in the triangle region of $x\le y$:
\begin{equation}
  \frac{\partial P_{\rvX\rvY}(x,y,t)}{\partial t} =
     D_a\frac{\partial^2 P_{\rvX\rvY}(x,y,t)}{\partial x^2} + 
     D_b\frac{\partial^2 P_{\rvX\rvY}(x,y,t)}{\partial y^2} - 
     V_a\frac{\partial P_{\rvX\rvY}(x,y,t)}{\partial x} +     
     \frac{F}{\eta_b}\frac{\partial P_{\rvX\rvY}(x,y,t)}{\partial y}.     
\end{equation}  

	The advantage of this version of the BR is its analytical
simplicity.  A coordinate transformation can be introduced:
\begin{equation}
    \rvD = \rvY-\rvX,   \hspace{0.5cm}
    \rvz = \frac{D_a \rvY + D_b \rvX}{D_a+D_b}    
\label{transf} 
\end{equation}
where $\rvD$ represents the gap between the tip of the actin
filament and the bacterium, $\rvz$ represents an averaged 
position of $\rvX$ and $\rvY$, we shall call it the {\it center
of mass} of the BR.  With this transformation, 
the two differential equations are decoupled:
\begin{equation}
   \frac{\partial P_{\rvD}(\Delta,t)}{\partial t} = (D_a+D_b)
   \frac{\partial^2 P_{\rvD}(\Delta,t)}{\partial \Delta^2} + 
        \left(V_a+\frac{F}{\eta_b}\right)
	\frac{\partial P_{\rvD}(\Delta,t)}{\partial \Delta},
   \hspace{0.5cm} (\Delta \ge 0)   
\label{eq4D}
\end{equation}
\begin{equation}
    \frac{\partial P_{\rvz}(z,t)}{\partial t} = 
    \frac{D_aD_b}{D_a+D_b}
   \left(\frac{\partial^2 P_{\rvz}(z,t)}{\partial z^2}\right) - 
     \frac{D_bV_a-D_aF/\eta}{D_a+D_b}   
   \left(\frac{\partial P_{\rvz}(z,t)}{\partial z}\right) 
   \hspace{0.3cm}  (-\infty < z < +\infty).
\label{eq4z}
\end{equation}
It can be immediately concluded from these two equations that
the gap $\rvD(t)$ approaches to its stationary, exponential
distribution (see Appendix)
\begin{equation}
        P_{\rvD}(\Delta) = \frac{V_a+F/\eta_b}{D_a+D_b} 
       e^{\frac{V_a+F/\eta_b}{D_a+D_b} \Delta}, \hspace{1cm}
        \Delta \ge 0.
\label{ddelta}
\end{equation}
$\rvz$ however increases steadily with an effective
diffusion constant $D_z$, 
$\frac{1}{D_z}=\frac{1}{D_a}+\frac{1}{D_b}$,
and a mean velocity $V_z=\frac{D_bV_a-D_aF/\eta_b}{D_a+D_b}$. 
This result can be understood in terms of Newtonian mechanics: 
the driving force from actin polymerization is 
$F_a=\eta_aV_a=k_BT V_a/D_a$, and the resistant force is $F$, 
and hence the net force on the BR is $F_z=F_a-F$ with the 
frictional coefficient the BR (center of mass) being 
$k_BT/D_z$.  Hence 
\begin{equation}
  V_z = \frac{D_zF_z}{k_BT} =
 \frac{D_aD_b}{D_a+D_b}\left(\frac{V_a}{D_a}-\frac{F}{k_BT}\right)
      = \frac{D_bV_a-D_aF/\eta_b}{D_a+D_b}.
\label{e4v}
\end{equation} 
The parameter $D_a$ and $V_a$ are defined in terms of
the $\alpha$, $\beta$, and $\delta$ in Eq. \ref{corresp}.
$\alpha$ is a pseudo-first order rate constant which
is proportional to the G-actin concentration $c_0$, as well
as the probability of the gap $\rvD$ being greater than
$\delta$.  Therefore, $\alpha$ is a function of
external force $F$; it can be determined in a self-consistent
manner by the transcendental equation:
\begin{equation}
     \alpha(F) = \alpha_0c_0 \int_\delta^{\infty}
                P_{\rvD}(s)ds
    =\alpha_0c_0\exp\left[-\frac{(\alpha-\beta)\delta+F/\eta_b}
                {(\alpha+\beta)\delta^2/2+D_b}\delta\right]
\label{eb}
\end{equation}
where $\alpha_0$ is the intrinsic, second-order rate constant for
polymerization.  We see that $V_z$ in Eq. \ref{e4v} is a linear
function of resistant force $F$ explicitly; however, nonlinearity
arises since $D_a$ and $V_a$ are implicit functions of the 
resistant force $F$, via $\alpha(F)$.  In other words, the 
resistant force $F$ slows down the BR by two different mechanisms: 
a linear Newtonian resistance and also a reduction in the rate of 
polymerization via a diminished gap.  Eq. \ref{eb} has the general
form $\alpha(F)\propto e^{-rF\delta/k_BT}$
with an entropic barrier.

	$V_z$ in Eq. \ref{e4v} is necessarily smaller than 
$V_a$, indicating that the bacterium 
retards the polymerization.  When $F=\eta_bD_bV_a/D_a$ = 
$\left(\frac{\eta_bD_b}{\delta}\right)
\frac{2(\alpha-\beta)}{\alpha+\beta}$, 
the bacterium completely stalls the polymerization.  
This yields the critical stalling force which agrees with
the well known result of Hill (1987) for an inert
object: $(k_BT/\delta)ln(\alpha/\beta)$.  Furthermore, 
if we note that $\eta_bD_b=k_BT$, then the rate of 
polymerization against a resisting force $F$ is at its maximal
$V_z=V_a-FD_a/k_BT$ when $D_b\rightarrow\infty$.  This is a result
of Peskin et al. (1993) who first elucidated the crucial role 
played by the fluctuating ``barrier'' in the filamental growth.  
In a more general context, the dynamic characteristics of 
the ``force transducer'' by which the resisting force is applied to 
the growing tip of the filament is an integral
part of the molecular process.\footnote{In the work 
of Hill (1981), this issue is not considered because of its 
quasi-thermodynamic approach.  The applied force was assumed to
have an instantaneous dynamic characteristics. From equilibrium 
thermodynamics one knows that polymerization
under resisting force $F$ has $\alpha(F)/\beta(F)=e^{-F\delta/k_BT}$.
So how does $F$ contribute to $\alpha(F)$ and $\beta(F)$ 
individually?  A splitting parameter $r$ is defined as
$\alpha(F) = \alpha(0)e^{-rF\delta/k_BT}$ and consequently
$\beta(F) = \beta(0)e^{(1-r)F\delta/k_BT}$, where $\alpha(0)$
and $\beta(0)$ are the $\alpha$ and $\beta$ in the main text. 
The rate of polymerization under force $F$, therefore, is 
\begin{equation}
            [\alpha(F)-\beta(F)]\delta = 
    \left(\alpha(0)e^{-rF\delta/k_BT} -\beta(0)e^{(1-r)F\delta/k_BT}
		\right)\delta. 
\label{split}
\end{equation}
With very small $\delta$, this gives 
$[\alpha(0)-\beta(0)]\delta - F\delta^2[r\ \alpha(0)+(1-r)\beta(0)]/k_BT$
which should be compared with 
$V_z$ = $(\alpha-\beta)\delta-F\delta^2(\alpha+\beta)/2k_BT$ 
from the main text.  This indicates that the BR has a splitting 
factor of $r=1/2$.  However, Hill's analysis does not have the 
contribution from the dynamic characteristics of the barrier, i.e., 
$D_b$. Eq. \ref{split} is the starting point of the recent work of
Kolomeisky and Fisher (2001).
}
	
	We now show that in the limit of $\delta \rightarrow 0$,
our $V_z$ from the continuous model is equivalent to the ratchet 
velocity derived by Peskin, Odell, and Oster (POO, 1993).  Note
that in the previous work the definition for the ratchet velocity 
was rather convoluted; The $V_z$ in the present model is more 
straightforward.  From Eq. \ref{corresp} we have $\alpha$ = 
$\frac{1}{2}\left(\frac{V_a}{\delta}+\frac{2D_a}{\delta^2}\right)$
and $\beta$ = 
$\frac{1}{2}\left(-\frac{V_a}{\delta}+\frac{2D_a}{\delta^2}\right)$.
Substituting these two expressions into 
\[  V_{poo} = \delta \left(\alpha\int_{\delta}^{\infty} 
   P_{\rvD}(\Delta) d\Delta - 
      \beta\int_0^{\infty} P_{\rvD}(\Delta)d\Delta
               \right)                    \]
we have
\[  \lim_{\delta \rightarrow 0} V_{poo} = V_a - \frac{D_a}{\delta}
        \int_0^{\delta} P_{\rvD}(\Delta)d\Delta
   = V_a - D_aP_{\Delta}(0)   
   = \frac{D_bV_a - D_aF/\eta_b}{D_a+D_b} = V_z.           \] 	
In the derivation we have used Eq. \ref{ddelta}. 
Note that the basic molecular parameters for polymerization, 
$\alpha$, $\beta$ and $\delta$ are actually contained in the 
parameter $D_a$ and $V_a$.  In the mathematical limit of $\delta
\rightarrow 0$, there is at the same time $\alpha$ and
$\beta$ $\rightarrow \infty$ such that $D_a$ and $V_a$ are
finite (Feller, 1957).

\vskip 0.5cm \noindent
{\bf 5. Discussion} 
\vskip 0.3cm \noindent

	Nanometre precision measurements on {\it L. monocytogenes} 
movement (Kuo and McGarth, 2000) have shown that the bacteria move 
with steps.  Considering there are many actin filaments in a bundle
which propels a bacterium, this observation indicates a synchronized 
filamental growth in the bundle. The synchronization is not 
inconsistent with a bundle of actin filaments propelling a
bacterium with sufficiently small Brownian movement (i.e., 
small $D_b$).  The significantly reduced Brownian motion is
indeed observed experimentally, both in the direction parallel and 
perpendicular to the actin growth.  

	There could be several explanations for the small $D_b$.  
a) Kuo and McGrath (2000) suggested an association between the 
bacterium and the actin structure, which leads to endorsing the 
two-dimensional BR with bending (Mogilner and Oster, 1996).  
b) It should be noted, however, that association-dissociation 
can also be introduced into the one-dimensional BR in the 
form of an attractive force between the actin and the bacterium;
thus a nonzero $F$ as function of $(y-x)$.  This, we suspect, 
will also lead to a reduced {\it apparent} $D_b$ on a longer 
time scale.  c) As we have pointed out, the $D_b$ of a living 
bacterium is not necessarily related to its physical size and 
frictional coefficient $\eta_b$.  A bacterium could have an 
internal mechanism, by utilizing its biochemical free energy, to 
localize itself near the tip of the actin filament with 
diminished Brownian motion.   Finally, all existing models on 
BR have only dealt with single filaments.  The continuous 
formalism we proposed here is in fact our initial step 
to extend the BR to a filamentous bundle.  All these topics
are currently under investigation.  
 
	In a special Science issue on Movement: Molecular to Robotic, 
two articles reviewed recent progress on force and motion generated 
on the molecular level by two completely different biological systems: 
motor protein movement and cytoskeletal filamental polymerization 
(Vale and Milligan, 2000; Mahadevan and Matsudaira, 2000).
Both systems can move against resistant force by utilizing chemical 
free energy.  In the abstract of the second article, it was stated 
``Not all biological movements are caused by molecular motors sliding 
along filaments or tubules.  Just as springs and ratchets can store 
or release energy and rectify motion in physical systems, their 
analogs can perform similar functions in biological systems.''  
While there has been much work done on motor proteins and protein 
polymerization in connection to various cellular phenomena such as 
motility, less has been discussed about the fundamental physical 
principles of these two processes.  It turns out that both molecular
processes have a single, unified mathematical model which accounts 
for their chemomechanical energy transduction.

	Theoretical formalism for motor proteins are now well 
established (see J\"{u}licher et al., 1997; Qian, 2000b, and 
references cited within).  Since the motion of a single motor 
protein is Brownian, it has to be characterized in terms of 
probability distribution.  The simplest model is that of
Huxley (1957).  This model corresponds to one on 
polymerization with nucleotide hydrolysis proposed by 
Dogterom and Leibler (1993).  Both models addressed the 
important issue of nucleotide hydrolysis, but neglected the 
stochastic nature in the movement of motor protein and actin 
polymerization, respectively.  Without the diffusion term,
such mathematical model is known as random evolution 
(Pinsky, 1991).

	With the ATP cap and hydrolysis, the model for the 
stochastic dynamics of actin polymerization will be precisely 
in the same class of the models for single motor proteins
(S.-D. Liang, G. Martinez, G.M. Odell, and H. Qian, work in 
progress).  The experimental measurements on both systems also 
proceed with parallel paths, as demonstrated by Dogterom and 
Yurke (1997), and more recently Kuo and McGrath (2000).  
Similar to the measurements on load-velocity curves for motor 
proteins, Dogterom and Yurke measured the velocity as a function 
of resistant force for single microtubules growing {\it in vitro}.  
Analysis of their data suggests that under the stalled (critical) 
condition, polymerization is in a nonequilibrium steady-state 
rather than a thermodynamic equilibrium (Kolomeisky and Fisher, 
2001; Hill, 1987).

	All these experimental evidences indicate that the class 
of BR model (or augmented Huxley model) is a fundamental mathematical 
model for chemomechanical energy transduction.  Recent work on the 
nonequilibrium statistical mechanics and thermodynamics of BR,
in the context for single macromolecules in aqueous solution, also
provided the mathematical model with a solid foundation in 
statistical physics of Boltzmann, Gibbs, and Onsager (Qian, 1998, 
2000b, 2001a,b,c).  The biological systems discussed in the 
two Science articles (Vale and Milligan, 2000; 
Mahadevan and Matsudaira, 2000) and the mechanistic, 
molecular models proposed are completely different. 
Yet they share fundamentally the same physiochemical principle 
which units both models in a quantitative fashion.  The 
mathematical model seems to capture the basic principle for 
molecular movements and forces in cell biology.

	To mathematical biologists, BR is a class of models which 
is based on a similar physical model but can have many different 
mathematical representations and different degree of 
approximations.  We have shown two such analyses in the present
work.  The essential feature of all the models can, and should,
be presented in terms of their MSD, which provides the BR,
in steady-state, with an effective diffusion constant($D_z$) and 
a mean velocity ($V_z$), both as functions of the resistant force.  
More subtle differences between models can be found in the transient
behavior.  The comparison between our analyses is summarized in 
the Table, in which the effective diffusion constant for the 
continuous model
\[
     D_z=\frac{(\alpha+\beta)\delta^2D_b/2}
	{D_b+(\alpha+\beta)\delta^2/2} \longrightarrow D_b
\]
when $\alpha\rightarrow\infty$, and the BR velocity 
\[
   V_z=\frac{D_b(\alpha-\beta)\delta-D_b(\alpha+\beta)\delta\omega/2}
       {D_b+(\alpha+\beta)\delta^2/2}
   \longrightarrow (D_b/\delta)(2-\omega)
\]
is a linear force-velocity relationship.  

\[
\begin{array}{l|cc}
	\hline\hline 
	&  D_z/D_b  & V_z\delta/D_b\\ \hline \\
   \textrm{discrete model} &   
      \frac{3e^{2\omega}-(10\omega-6)e^{\omega}
	+\omega^2-2\omega-9}{(e^{\omega}-1-\omega)^3}\omega^2
			&\frac{\omega^2}{e^{\omega}-1-\omega} \\
\\[5pt]
   \textrm{continuous model}& 1 & 2-\omega \\[5pt]
	\hline\hline
\end{array}
\]

	It is seen that in the continuous model, the effective
diffusion constant $D_z$ is always less than the $D_b$, while in the 
discrete model, the $D_z=\sigma^2/2$ is a function
of the resistant force.  When the force is small, 
$D_z$ can in fact be greater than $D_b$.  This is a type of
facilitated diffusion. 

	It is important to point out that the results from 
our discrete analysis is invalid when the resistant $F$ is 
sufficiently large, when the polymerization is near its stalling 
force.  This is due to the assumption of infinite large $\alpha$.  
This explains why there is no critical force in Fig. 3, at which 
the velocity $v=0$.  The more realistic model with finite $\alpha$ and 
$\beta$ does lead to a finite, positive stalling force
(Peskin et al., 1993; Kolomeisky and Fisher, 2001).  
The valid regime for our discrete model is a rapid growing 
actin filament with the bacteria viscous drag being the 
limiting factor in the overall BR movement.

	Finally, it is worth pointing out that mechanical studies 
of cellular properties and functions can be approximately 
classified as for passive and active materials.  The former can 
be understood in terms of the theories of viscoelasticity and 
polymer dynamics, see Qian (2000a) for a general approach to the 
problem.  Materials with chemomechanical energy transduction are 
active.  The fundamental difference is the nucleotide hydrolysis 
which leads to an irreversible thermodynamic nonequilibrium 
steady-state, with heat dissipation (Qian, 2001c), in the latter 
rather than the usual equilibrium.  Thus, the BR model is also 
a natural generalization of the standard polymer theory for 
passive materials (Doi and Edward, 1986) to active materials for 
which T.L. Hill (1987) has coined the term ``steady-state polymer''.  
There is a continuous intellectual thread in all these 
mathematical theories.

\vskip 0.5cm \noindent
{\bf 6. Acknowledgement} 
\vskip 0.3cm \noindent

	I thank Scot Kuo, Gilbert Martinez, and Gary Odell for many 
helpful discussions, Elliot Elson and  Charles Peskin for helpful 
comments on the manuscript.

\vskip 0.5cm \noindent
{\bf 8. Appendix} 
\vskip 0.3cm \noindent

\vskip 0.3cm
{\bf\em Diffusion with Drift in Semi-infinite Space with Noflux 
Boundary.}  To understand the dynamics of the gap in the continuous 
model, one needs to solve the time-dependent diffusion equation, 
Eq. \ref{eq4D}.  In nondimensionalized form:
\begin{equation}
        u_t = u_{xx} + \omega u_x, \hspace{0.3cm} (x\ge 0),
\label{seminf}
\end{equation}
with boundary condition $u_x+\omega u$ = 0 at $x=0$ and $x=\infty$.
Amazingly, classic texts on diffusion (Carslaw and Jaeger, 1959; 
Crank, 1975) did not give an explicit solution to the problem. 
Becasue of its central importance in the theory of BR, we give 
some explicit results below.  

The eigenfunction of the problem associated with the eigenvalue
$\mu(z)$ = $-\frac{1}{4}(\omega^2+z^2)$ is
\begin{equation}
      u(x,t;z) = \frac{1}{\sqrt{\pi(\omega^2+z^2)}}\left[
		z\cos\left(\frac{zx}{2}\right)
	-\omega\sin\left(\frac{zx}{2}\right)\right]
	e^{-\frac{\omega x}{2}+\mu(z)t}, \hspace{0.3cm}
							(z\ge 0),
\label{eigenf}
\end{equation}
and for $\mu=0$, $\sqrt{\omega}e^{-\omega x}$. Note there is a 
gap in $\mu$ between $\mu=0$ and the continuous spectrum
$\mu\le-\omega^2/4$. The Sturm-Liouville eigenvalue problem 
has a complete orthonormal set 
\[  \int_0^{\infty}  u(x,0;z)u(x,0,z')e^{\omega x}dx 
		=  \delta (z-z').
\]
Therefore, the solution to Eq. \ref{seminf} with initial data 
$\delta(x)$ is
\begin{equation}
      u(x,t) =\omega e^{-\omega x} +  
		e^{-\frac{\omega x}{2}-\frac{\omega^2t}{4}}
         \int_0^{\infty} \frac{zdz}{\pi(\omega^2+z^2)}
      \left[z\cos\left(\frac{zx}{2}\right)
	-\omega\sin\left(\frac{zx}{2}\right)\right]
	e^{-z^2t/4},
\label{udis}
\end{equation}
which approaches to the exponential distribution $\omega e^{-\omega t}$
when $t\rightarrow\infty$.  From Eq. \ref{udis} we have
\[    \int_0^{\infty} u(x,t)dx = 1,
\]
\begin{eqnarray}
     \int_0^{\infty}xu(x,t)dx &=& \frac{1}{\omega} -
		e^{-\frac{\omega^2t}{4}}
         \int_0^{\infty} \frac{4z^2e^{-z^2t/4}}
			{\pi(\omega^2+z^2)^2}dz
\nonumber\\
    &=&\frac{1}{\omega}\left[(1+2\tau)erf\left(\sqrt{\tau}\right)
		-2\tau+2\sqrt{\frac{\tau}{\pi}}e^{-\tau} \right],
\label{gapf}
\end{eqnarray}
where $\tau=\omega^2t/4$. The curve in the square bracket is a
universal curve, we shall denote it by $1-gap(\tau)$.  See Fig. 
6.  It is interesting to point out that there is a sharp 
transition between an unlimited growth of $x$ for $\omega<0$ and 
a stationary state for $x$ when $\omega>0$.  This mathematical 
result is similar to that of Dogterom and Leibler (1993).  
Finally,
\[
      \int_0^{\infty}x^2u(x,t)dx = \frac{8}{\omega^2}
      \int_0^{\tau} gap(s)ds.
\]

\vskip 0.3cm
{\bf\em The Dynamics of Gap $\rvD(t)$ in the Discrete Model.}  
Eq. \ref{eigen} has a second boundary condition $u=0$ at $x=\delta$,
which corresponds to nondimensionalized Eq. \ref{seminf} with 
$0\le x\le 1$ and $u(1)=0$.  It has a set of discrete eigevalues.  
The eigenfunctions to the nondimensionalized Eq. \ref{eigen} with 
eigenvalue $\mu(z)$ = $-\frac{1}{4}(\omega^2+z^2)$ $\le$ 0, 
are still given in Eq. \ref{eigenf}, but the $z$'s are now the
discrete roots of the transcendental equation 
$\cos z $ = $\frac{\omega^2-z^2}{\omega^2+z^2}$.  When $\omega<2$,
the equation for $z$ has only real roots; hence 
the largest eigenvalues is $\mu<-\omega^2/4$.
If, however, $\omega>2$, then there is a pair of imaginary 
$\pm iz^*$, $|z^*|<\omega$. Then  the largest eigenvalue is 
$\mu=-\left(\omega^2-(z^*)^2\right)/4$.
Fig. 4 shows how the largest eigenvalue (smallest
in magnitude) changes as functions of $\omega$.  It is seen
that the largest eigenvalue can be well represented by $1/E[\rvT]$
\begin{equation}
       \mu_1 \approx -\frac{\omega^2}{e^{\omega}-1-\omega}
	   \hspace{0.3cm} \textrm{or} \hspace{0.3cm}
       z_1 = \omega\sqrt{\frac{4}{e^{\omega}-1-\omega}-1}.
\end{equation}
The solution to the time-dependent Eq. \ref{eigen} with initial data 
$\delta(x)$ can be obtained in terms of the $u(x,t;z)$'s in
Eq. \ref{eigenf}, and approximated by the first term: 
\begin{equation}
      u(x,t) \approx \frac{z_1}{\pi(\omega^2+z_1^2)} \left[
        z_1\cos\left(\frac{z_1x}{2}\right) - \omega
	\sin\left(\frac{z_1x}{2}\right) \right] 
	\exp\left(-\frac{\omega}{2}x-
		\frac{\omega^2+z_1^2}{4}t
		\right).
\end{equation}

	Therefore,
\[
     \int_0^1 u(x,t)dx = \frac{2z_1}{\pi(\omega^2+z_1^2)}
		e^{-\frac{\omega}{2}-\frac{\omega^2+z_1^2}{4}t}
		\sin\left(\frac{z_1}{2}\right),
\]
which tends to zero because the exit probability.  However, 
the remaining probability for $\rvD(t)$ quickly approaches to a 
quasi-stationary distribution 
\begin{eqnarray*}
    f_{\rvD}(x) &=& \frac{z_1\cos\left(\frac{z_1x}{2}\right) - \omega
	\sin\left(\frac{z_1x}{2}\right)}
	{2\sin\left(\frac{z_1}{2}\right)}
	\exp\left(-\frac{\omega}{2}(x-1)\right)
\\
      &=& \frac{\omega^2+z_1^2}{2z_1}
		\sin\left(\frac{z_1}{2}(1-x)\right)
            	\exp\left(\frac{\omega}{2}(1-x)\right).
\end{eqnarray*}
For large $\omega$, this distribution approaches to the exponential
distribution $\omega e^{-\omega x}$ as expected (data not shown).

\newpage

\begin{figure}[h]
\[
\psfig{figure=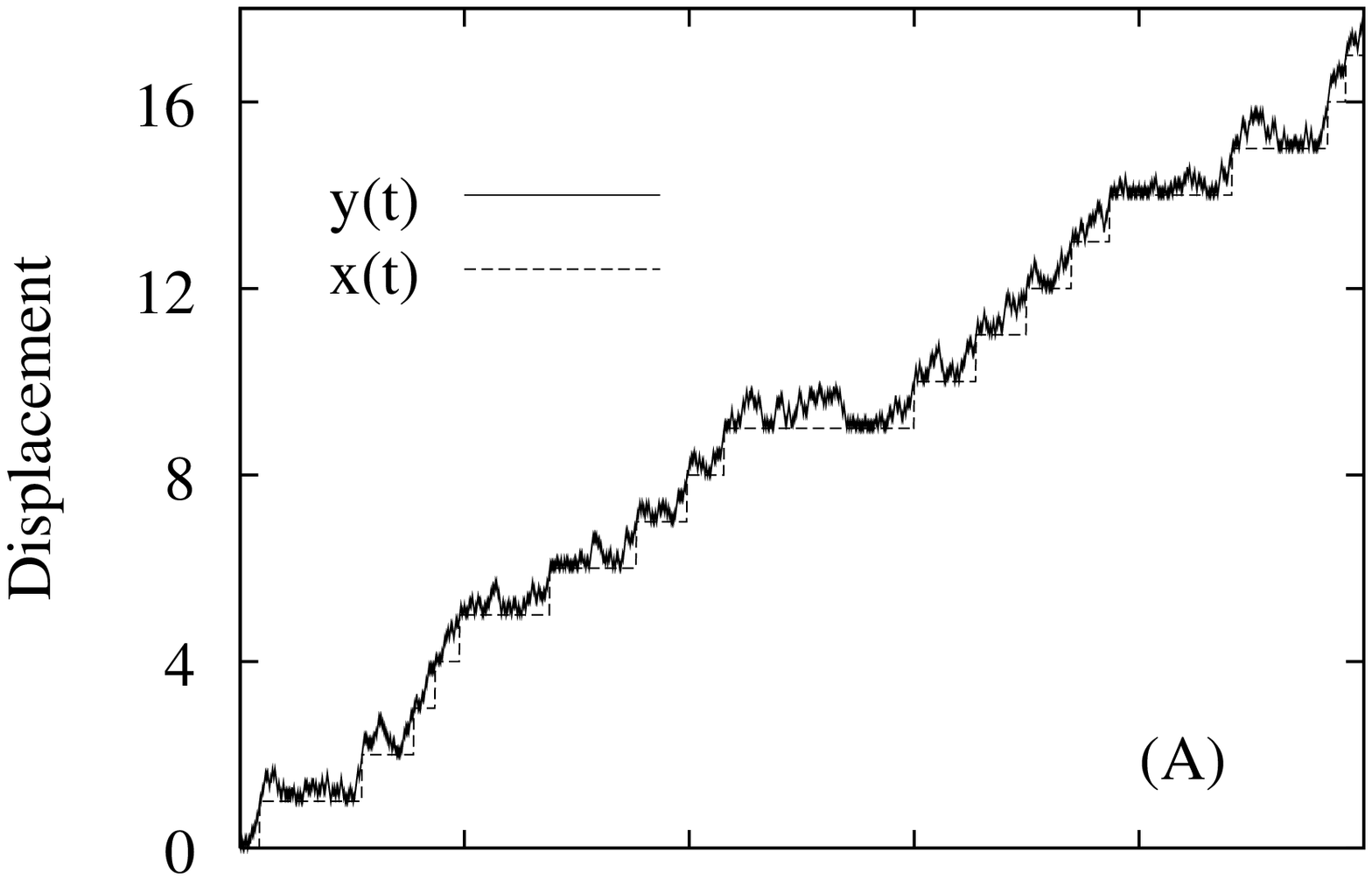,%
width=5.1in,height=3.75in,%
bbllx=2.5in,bblly=0.25in,%
bburx=8.5in,bbury=8.in,%
angle=-90}
\]
\vskip -3.2cm
\[
\psfig{figure=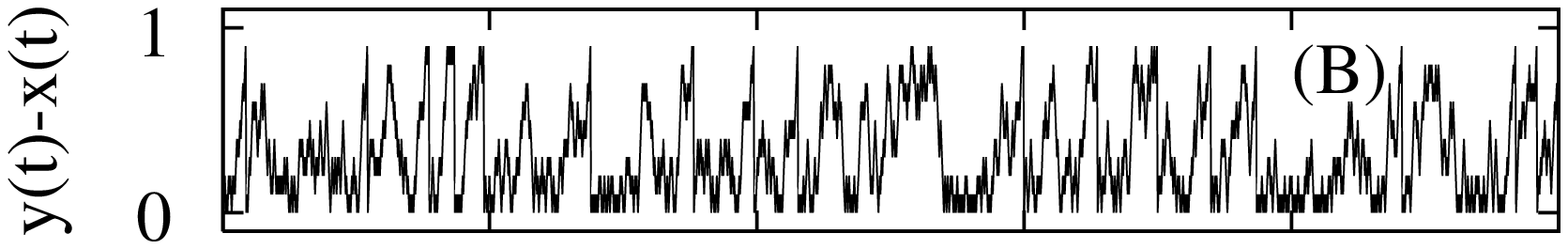,%
width=5.1in,height=2in,%
bbllx=5.5in,bblly=0.0in,%
bburx=8.5in,bbury=8.in,%
angle=-90}
\]
\vskip -2.3cm
\[
\psfig{figure=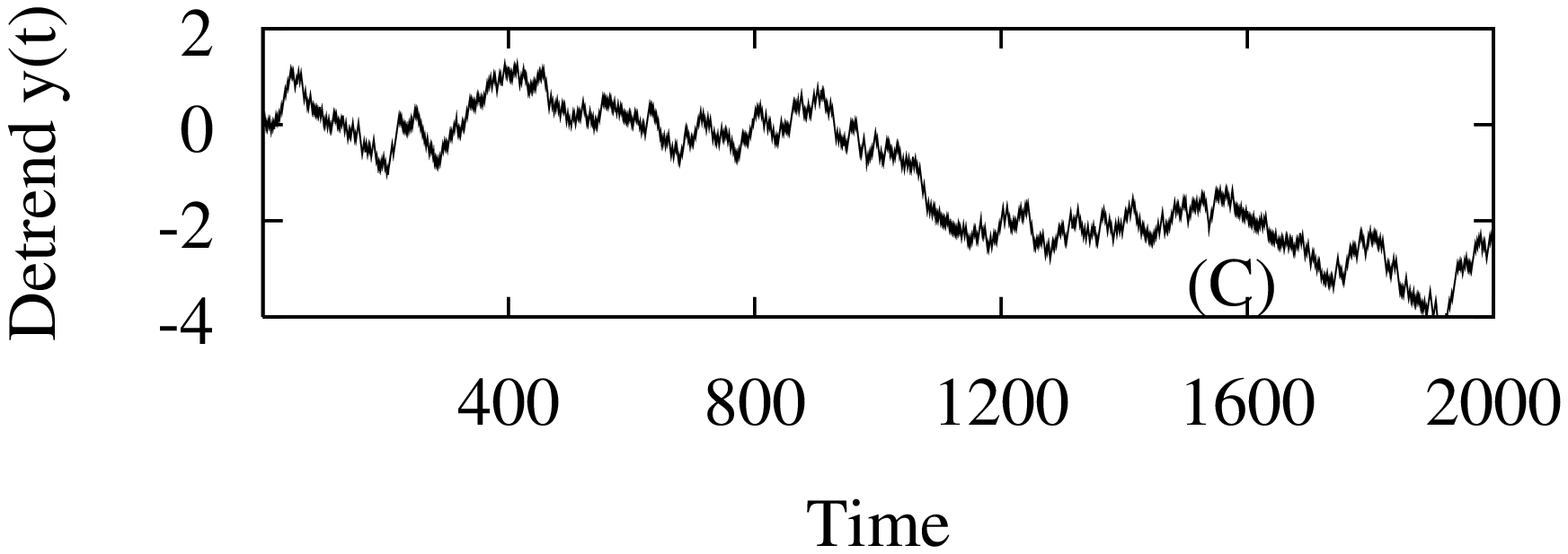,%
width=5.1in,height=2.in,%
bbllx=5.5in,bblly=0.25in,%
bburx=8.5in,bbury=8.in,%
angle=-90}
\]
\caption{A set of examples, from Monte Carlo simulations, for the 
stochastic trajectories of $\rvY(t)$, the movement of the bacterium, 
$\rvX(t)$, the movement of the tip of the actin filament, 
$\rvD(t)\triangleq\rvY(t)-\rvX(t)$, 
the gap, and $\hat{\rvY}(t)$, the detrended $\rvY(t)$. Among the four 
types of data, only the $\rvD(t)$ approaches stationarity. 
$\hat{\rvY}(t)$ has zero expectation but with a linear MSD, a 
characteristic of Brownian motion without drift. Both $\rvX(t)$ and 
$\rvY(t)$ show the typical diffusion with a drift (Qian et al., 1990). 
}
\end{figure}

\pagebreak

\begin{figure}[h]
\[
\psfig{figure=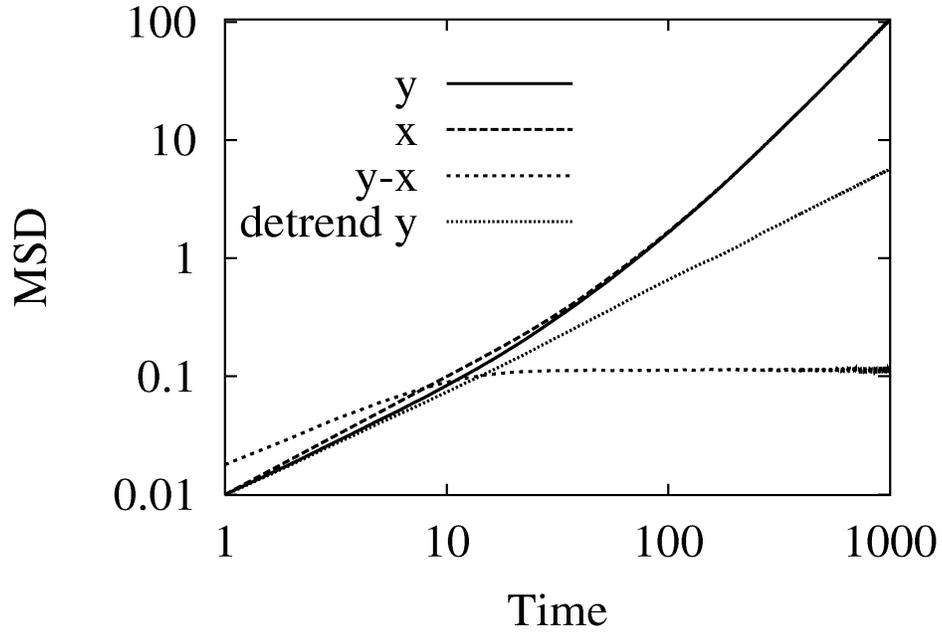,%
width=5.in,height=4.in,%
bbllx=2.5in,bblly=0.75in,%
bburx=8.5in,bbury=8.in,%
angle=-90}
\]
\caption{The MSD calculated for the four types of data in Fig. 1.
$\rvY(t)$, the movement of the bacterium, $\rvX(t)$, the movement of 
the tip of the actin filament, $\rvD(t)$, the gap, and 
the detrend $\hat{\rvY}(t)$. As expected, after a brief period of
time, the $\rvY(t)$ and $\rvX(t)$ are almost indistinguishable; the 
gap between them quickly reaches stationarity. The detrend 
$\hat{\rvY}(t)$ shows a linear MSD, as observed in the experiments 
(Kuo and McGarth, 2000).}
\end{figure}

\pagebreak

\begin{figure}[h]
\[
\psfig{figure=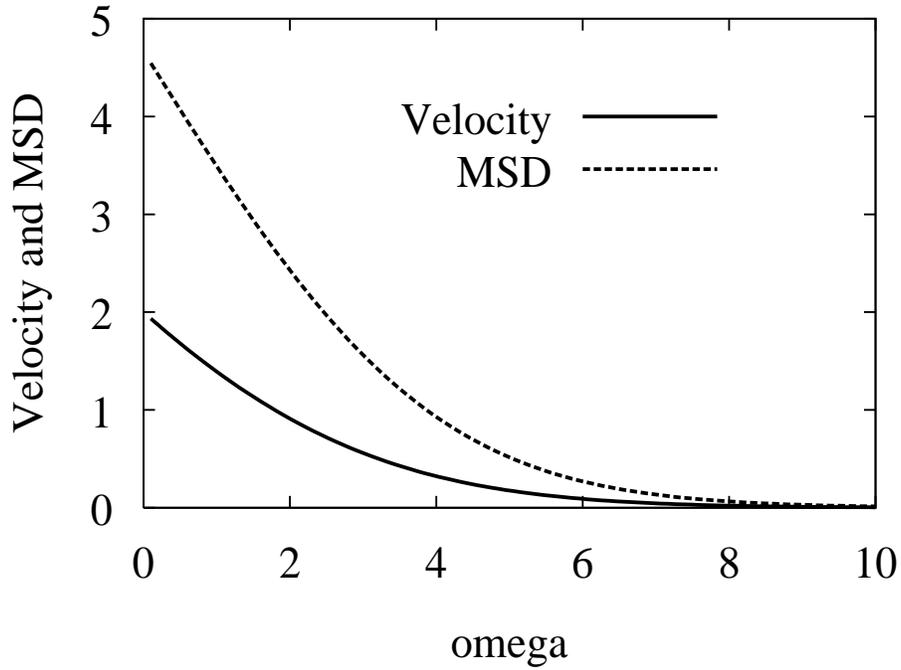,%
width=5.in,height=4.25in,%
bbllx=2.5in,bblly=0.75in,%
bburx=8.5in,bbury=8.in,%
angle=-90}
\]
\caption{The velocity $v\delta/D_b$ and MSD $\sigma^2/D_b$, 
nondimensionalized, as functions of the resistant force on bacterium, 
$\omega = F\delta/\eta_bD_b$.   In the SPT experiments, the velocity 
can be obtained as the quadratic term in the MSD of $\rvY(t)$, 
the $\sigma^2$ can be obtained either as the linear term in 
the MSD of $\rvY(t)$, or the MSD of the detrend $\hat{\rvY}(t)$.  
The velocity $v$ is given in Eq. \ref{BRv} and the $\sigma$ 
is given in Eq. \ref{sig}.}  
\end{figure}

\pagebreak

\begin{figure}[h]
\[
\psfig{figure=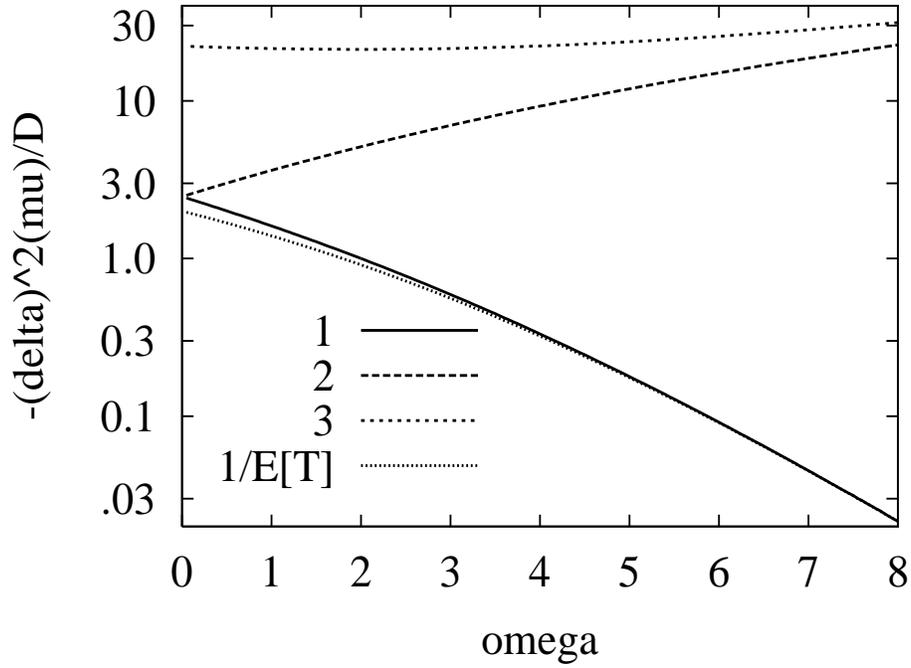,%
width=5.in,height=4.25in,%
bbllx=2.5in,bblly=0.75in,%
bburx=8.5in,bbury=8.in,%
angle=-90}
\]
\caption{Numerical computation for the three smallest eigenvalues 
(in magnitude, all eigenvalues are negative) of Eq. \ref{eigen},
$\left|\mu\delta^2/D_b\right|$, as function of the 
resistant force $\omega$.  These are the most relevant modes in the 
relaxation (and correlation function) of the gap, $\rvD(t)$, approaching 
to stationarity. The smallest eigenvalue corresponds to the
exit time, shown in the figure (labeled $1/E[\rvT]$).}
\end{figure}

\pagebreak

\begin{figure}[h]
\[
\psfig{figure=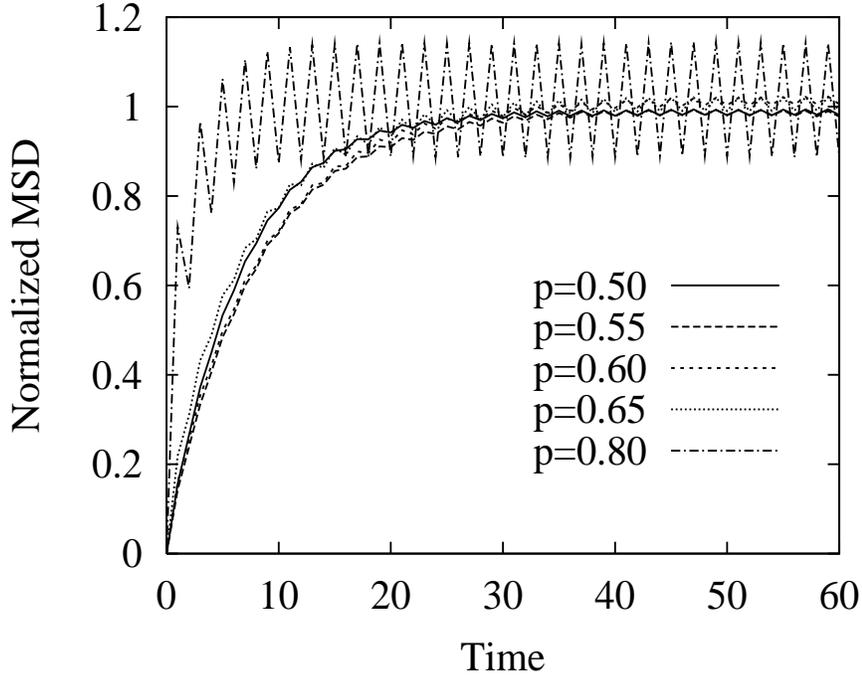,%
width=5.in,height=4.25in,%
bbllx=2.5in,bblly=0.75in,%
bburx=8.5in,bbury=8.in,%
angle=-90}
\]
\caption{The normalized MSD for the $\rvD(t)$, from Monte Carlo simulation, 
for different resistant force $F$ which is related to the probability 
($p$) shown in the figure: $F/(\eta_b\delta)$ = $0.1(2p-1)$.  A standard 
MSD for a stationary process is directly related to its time 
correlation function $2\left(E[\rvD^2(t)]-E[\rvD(t)\rvD(0)]\right)$, 
with its asymptote 
being the $2Var[\rvD]$ when $t\rightarrow\infty$. In the simulations,
the diffusion constant is $D_b/\delta^2$ = 0.005.  $0\le\rvD(t)\le 1$; 
the stationary $E[\rvD]$ = 0.33, 0.27, 0.21, 0.16, and 0.08 for 
$p=0.5-0.8$ respectively; the corresponding relative variances 
$Var[\rvD]/E^2[\rvD]$ = 0.53, 0.64, 0.79, 0.91, and 0.96.  The
relative variance for an exponential distribution is 1.}
\end{figure}

\pagebreak

\begin{figure}[h]
\[
\psfig{figure=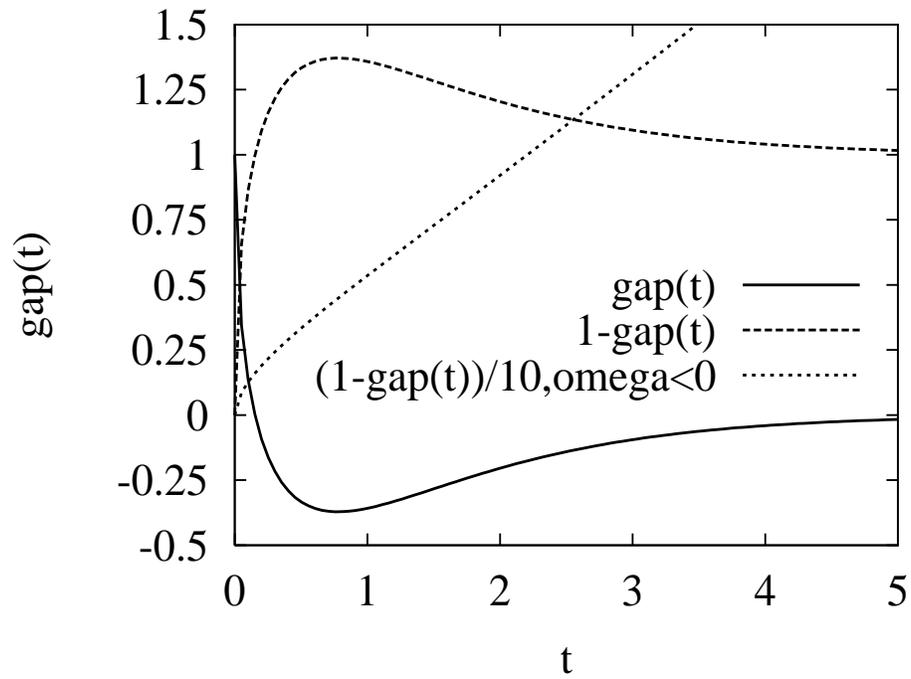,%
width=5.in,height=4.25in,%
bbllx=2.5in,bblly=0.75in,%
bburx=8.5in,bbury=8.in,%
angle=-90}
\]
\caption{The function $gap(t)$, defined in Eq. \ref{gapf}, is the 
time course for the the mean gap size to approach to its stationarity. 
For $\omega>0$, the mean gap size approaches to a finite size,
while for $\omega<0$, the mean gap size grows without bound.} 
\end{figure}

\end{document}